\documentclass[aps, prl, twocolumn,nofootinbib, superscriptaddress,longbibliography]{revtex4-2}
\usepackage{graphicx}
\usepackage{amsfonts}
\usepackage{float}
\usepackage{amsmath}
\usepackage{latexsym}
\usepackage{bm, bbold}
\usepackage{amsmath,amssymb}
\usepackage{hyperref}
\usepackage{graphicx}
\usepackage[export]{adjustbox}
\usepackage[T1]{fontenc}
\usepackage{times}
\usepackage{xcolor}
\usepackage{lineno}

\begin{document}

\onecolumngrid
\begin{center}
  \textbf{\large Ultrafast Many-Body Dynamics in an Ultracold Rydberg-Excited Atomic Mott Insulator}\\[0.2cm]
\end{center}

\begin{center}
    {\normalsize
    \footnotetext{These authors contributed equally to the work.}}
    V. Bharti$^{1,}$\footnotemark[\value{footnote}],
    S. Sugawa$^{1,2,}$\footnotemark[\value{footnote}]$^{,}$\footnote[2]{sugawa@ims.ac.jp},
    M. Mizoguchi$^{1,}$\footnotemark[\value{footnote}],
    M. Kunimi$^{1}$,
    Y. Zhang$^{1,3}$,\\
    S. de L\'es\'eleuc$^{1,2}$,
    T. Tomita$^{1}$,
    T. Franz$^{4}$,
    M. Weidem\"uller$^{4}$,
    K. Ohmori$^{1,2,}$\footnote[3]{ohmori@ims.ac.jp}
    \\[.2cm]
    {\small
    $^{1}$\textit{
    Institute for Molecular Science, National Institutes of Natural Sciences, Okazaki 444-8585, Japan}\\
    $^{2}$\textit{
    SOKENDAI (The Graduate University for Advanced Studies), Okazaki 444-8585, Japan}\\
    $^{3}$\textit{
    College of Physics and Electronic Engineering, and Collaborative Innovation\\ Center of Extreme Optics, Shanxi University, Taiyuan, Shanxi 030006, China}\\
    $^{4}$\textit{
    Physikalisches Institut, Universit\"at Heidelberg, Im Neuenheimer Feld 226, 69120 Heidelberg, Germany}\\
  }
\end{center}
\vspace*{-0.3cm}

\date{\today}

\begin{abstract}
    We report the observation and control of ultrafast non-equilibrium many-body electron dynamics in Rydberg-excited spatially-ordered ultracold atoms created from a three-dimensional unity-filling atomic Mott insulator.
    By implementing time-domain Ramsey interferometry with attosecond precision in our Rydberg atomic system, we observe picosecond-scale ultrafast many-body dynamics that is essentially governed by the emergence and evolution of many-body correlations between long-range interacting atoms in an optical lattice. 
    We analyze our observations with different theoretical approaches and find that quantum fluctuations have to be included beyond semi-classical descriptions to describe the observed dynamics.
    Our Rydberg lattice platform combined with an ultrafast approach, which is robust against environmental noises, opens the door for simulating strongly-correlated electron dynamics by long-range van der Waals interaction and resonant dipole-dipole interaction to the charge-overlapping regime in synthetic ultracold atomic crystals.
\end{abstract}

\maketitle

Well-controlled isolated quantum systems have become an essential experimental platform to address quantum many-body problems that are inaccessible with classical computers~\cite{Georgescu2014_RevModPhys}.
Of particular interest is the understanding of the emergence of many-body correlations and entanglement that arise from long-range interaction between particles.
Quantum spin models with long-range interaction have been realized in
various systems including trapped ions~\cite{Kim2010_Ion, Smith2016_IonMBL, Bohnet2016_2Dion}, polar molecules~\cite{Micheli2006_polar, Yan2013_RbK}, and magnetic atoms~\cite{Lepoutre2019, Patscheider2020_ErXXZ}. 
Recently, Rydberg gases have emerged as a promising system to study many-body problems due to their high controllability and tunable long-range interaction~\cite{Browaeys2020_NphysReview, deLeseleuc_SPTphase, Bluvstein2021_Scar, Omran2019_Cat}.
Spin models have been implemented in one, two, or three dimensions with spatially ordered and disordered atomic systems~\cite{Takei2016_NCommun, Zeiher2016_2dIsing, Zeiher2017_1dDressedIsing, Bernien2017_51simulator, Schauss_2dIsingCrystal, Labuhn2016_2dIsing, Lienhard2018_PRX, Signoles2021_3dGlassy, Orioli2018_spin_relaxation, GuardadoSanchez2018PRX, Ebadi2021_256Ryd, Scholl2021_196Ryd, Geier2021_Floquet}.
These systems offer opportunities to address fundamental questions on many-body spin physics, such as the understanding of the origin of quantum magnetism and the emergence of many-body correlations and entanglement during the out-of-equilibrium dynamics. 

While many experimental studies on many-body dynamics using Rydberg atoms have worked with continuous-wave (CW) laser excitation, ultrashort laser pulses excitation is expected to offer unique possibilities to explore novel ultrafast strongly-correlated electron dynamics. With CW laser excitation, simultaneous excitation of atomic pairs at a short atomic distance typically below several microns is inhibited due to the Rydberg blockade effect~\cite{Singer2004_PRL,Tong2004_PRL,Heidemann2007_PRL, Gaetan2009, Urban2009}.
Facilitated excitation with frequency-detuned CW laser has been developed to reach shorter atomic distance, however, it gives strong constraints~\cite{Urvoy2015, Schempp2014, Malossi2014, Viteau2012, Amthor2007}. 
Broadband laser excitation with an ultrashort pulse can directly circumvent the blockade effect even at the nearest-neighbor (NN) distance in optical lattices~\cite{Mizoguchi2020_PRL}.
This leads to the highest Rydberg density with GHz-scale interaction, three orders of magnitude larger than the CW approach~\cite{Ebadi2021_256Ryd, Scholl2021_196Ryd, Takei2016_NCommun}.
The interaction timescale is many orders of magnitude shorter than the radiative lifetimes of the Rydberg states and the timescale set by environmental noises (both shot-to-shot and dynamical), thermal motions, and laser phase noises~\cite{deLeseleuc2018_imperfect,Carter2013_ChipRyd, Baluktsian2012_PfauPRL}, thus making our ultrafast approach advantageous for exploring long-time out-of-equilibrium many-body dynamics.
A previous experiment using a spatially-disordered ensemble of micro-kelvin Rubidium (Rb) atoms in an optical dipole trap revealed the ultrafast dynamics of many-body electron coherence between the ground and high-lying electronic (Rydberg) states. 
More than 40 particles were shown to be correlated in several hundreds of picoseconds~\cite{Takei2016_NCommun}.
This work opened a new direction in the field in which an ultrafast approach is combined with an ultracold atomic system to explore many-body electron dynamics. 
The extension to spatially-ordered atomic arrays, which has been elusive so far, could further shed light on novel many-body coherence phenomena on ultrafast timescales.

In this Letter, we report the first observation of ultrafast non-equilibrium many-body electron dynamics in Rydberg-excited atoms created from a three-dimensional unity-filling atomic Mott insulator (MI). 
By employing time-domain Ramsey interferometry~\cite{Ohmori2003}, we succeeded in observing many-body coherence dynamics in the picosecond timescale that arise from the long-range interaction between Rydberg atoms. 
We analyze our result with a hierarchy of theoretical models and gain insights into the buildup of the long-range quantum correlations in our ultrafast dynamics.

\begin{figure}
    \centering
    \includegraphics[width=86mm]{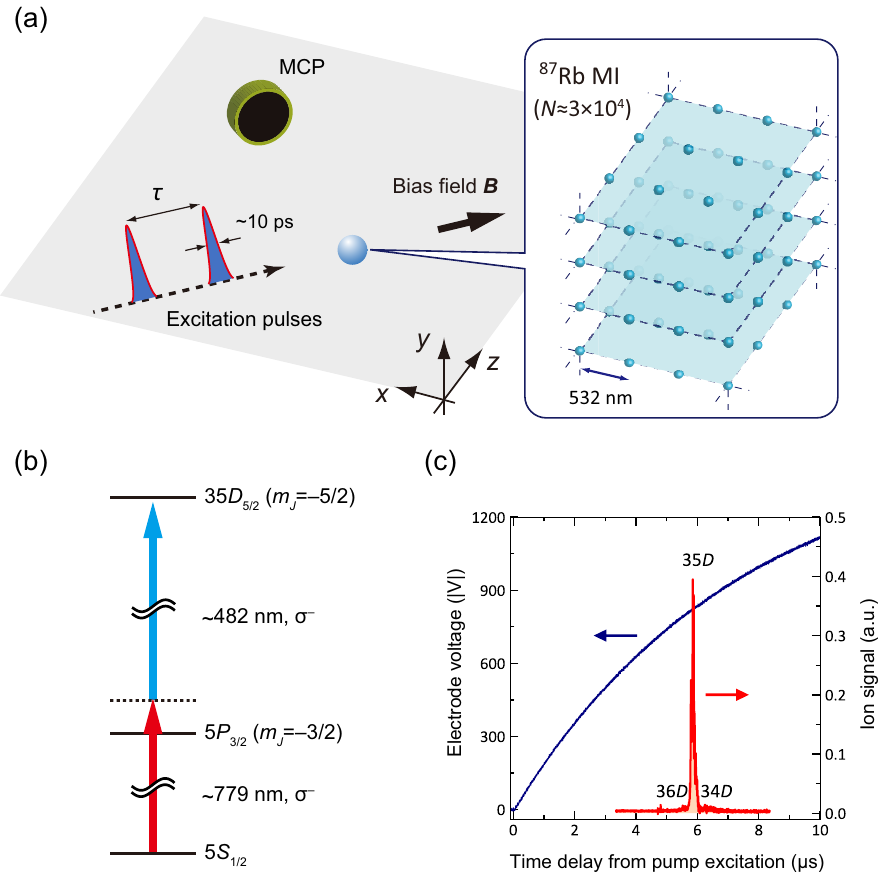}
    \caption{
    (a) Experimental schematic. 
    The atoms are irradiated with a pair of picosecond excitation pulses with tunable time-delay $\tau$ during which the system evolves. The atoms in the Rydberg state are detected as ions by a MCP following field-ionization.
    (b) The atoms in the ground state are coherently excited to the $35D$ state by two-photon laser excitation with circularly polarized blue and IR laser pulses propagating along the bias magnetic field direction.
    (c) TOF spectrum of Rydberg states following field-ionization. The ion signal detected at the MCP (red) is shown with the electrode voltage for field-ionization (blue). 
    The Rydberg states are resolved by the different arrival times in the TOF. The measurement is performed with a large micro-kelvin temperature atomic cloud to avoid prompt ionization.
    }
    \label{fig1}
\end{figure}

\emph{Experimental setup} 
Our experimental system is schematically shown in Fig.~1(a).
Our experiment begins with $N\approx 3 \times 10^{4}$ $^{87}$Rb atoms in a three-dimensional optical lattice at a lattice depth of $\sim 20E_\mathrm{R}$ forming a unity-filling MI~\cite{Mizoguchi2020_PRL}.
Here $E_\mathrm{R} = h^{2}/2m_\mathrm{Rb}\lambda^{2}$ is the recoil energy of the lattice laser operating at $\lambda = 1064\,$nm, and $m_\mathrm{Rb}$ is the $^{87}$Rb mass.
The optical lattice is formed by superimposing three orthogonal standing-waves of light with a spatial periodicity of $\lambda/2$, thereby creating a trap potential in a cubic lattice geometry with a lattice constant of $a_\mathrm{lat} = 532\,$nm.
The Rb atoms in the MI act as a nearly defect-free three-dimensional single-atom array for our experiment.

The atoms which are initially in the hyperfine ground state, $5S_{1/2}$, $|F=2, m_{F}=-2\rangle$ are coupled to a Rydberg state via two-photon optical transition using broadband picosecond laser pulses with their wavelengths tuned to $\sim 779\,$nm and $\sim 482\,$nm (Fig.~1(b)).
The polarizations of the laser pulses are both set to $\sigma^{-}$ so that only $|\nu D_{5/2}$, $m_J=-5/2\rangle$ can be populated from an optical selection rule. Here, $\nu$ is the principal quantum number.
In the present study, $\nu=35$ is chosen, and the center laser frequencies are tuned to its two-photon resonance.
The excitation bandwidth (HWHM: $\sim 85\,$GHz) is set to have a negligible population on the nearby $36D_{5/2}$ and $34D_{5/2}$ states, which locate $\sim 165\,$GHz above and $\sim 181\,$GHz below the $35D_{5/2}$ state, respectively as confirmed in the time-of-flight (TOF) spectrum of the Rydberg population after a pulse excitation (Fig.~1(c)).
With a single pump excitation pulse, we coherently excite $p_\mathrm{e} \sim 5.6\%$ of the initial ground state atoms to the Rydberg state by $\sim 50\,$nJ of the IR laser pulse and $\sim 565\,$nJ of the blue laser pulse.

\begin{figure}
    \centering
    \includegraphics[width=86mm]{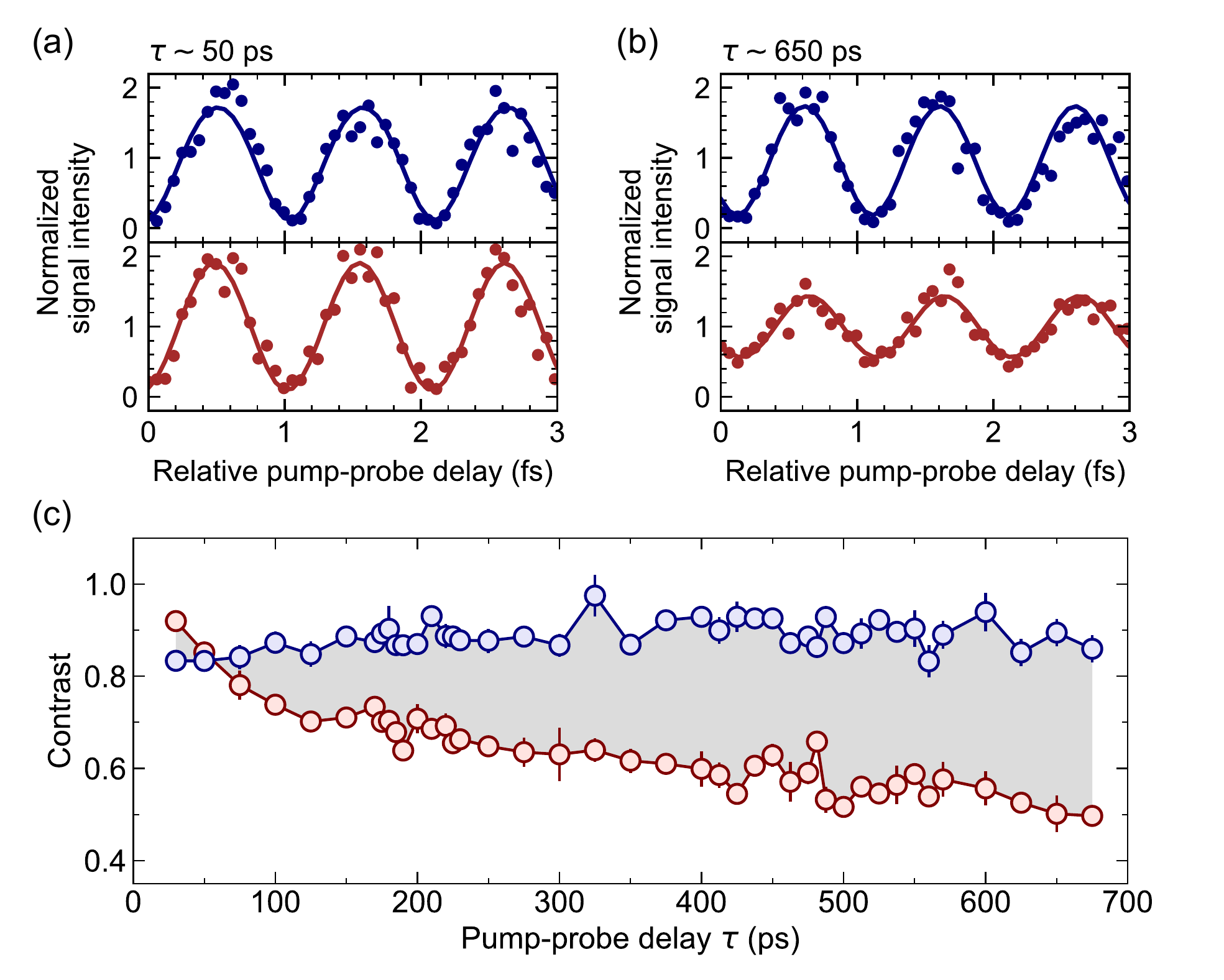}
    \caption{
    (a),(b) Time-domain Ramsey interferograms for MI (red) and reference atomic sample (blue) are shown for two different delays ($\tau\sim50\,$ps (a) and $\tau\sim650\,$ps (b)).
    Solid lines are theory fit to the data.
    (c) The contrast of the Ramsey signal for MI (red) and reference atomic sample (blue) with $p_\mathrm{e} = 5.6(2)\%$. 
    Error bars represent s.e.m.
    }
    \label{fig2}
\end{figure}

\emph{Ramsey interferogram}
We measure many-body electron dynamics by time-domain Ramsey interferometry with a pair of excitation laser pulses~\cite{Liu2018_Ramsey,Katsuki2009_Spatiotemporal}.
The first pump pulse (time origin) excites the ground state atoms to the Rydberg state creating a superposition between $5S_{1/2}$ and $35D_{5/2}$ states. The system undergoes many-body dynamics that originate from long-range and anisotropic interaction between Rydberg atoms until the second probe pulse irradiates the atoms.
The time-delay between the pump pulse and the probe pulse is tuned with attosecond precision using an optical delay line interferometer~\cite{Katsuki2009_Spatiotemporal}.
The two pulses have identical pulse energy and property.

Ramsey oscillations can be observed by measuring the final Rydberg population as a function of the time-delay. In the absence of interaction and decoherence, the Rydberg population $P_\mathrm{e}(\tau)$
temporally oscillates as~\cite{Liu2018_Ramsey}
\begin{equation}
    p_\mathrm{e}(\tau) = 2 p_\mathrm{e} p_\mathrm{g}[1 + \cos( E_\mathrm{eg}\tau/\hbar + \phi)],    
    \end{equation}
where $p_\mathrm{e}$ ($p_\mathrm{g}$) is the Rydberg (ground) state population after the pump excitation pulse, $E_\mathrm{eg}$ is the energy difference between ground and Rydberg states and $\phi$ is a phase.
After irradiating the probe pulse, we apply a short electric field pulse to ionize the Rydberg atoms, which are finally detected at a micro-channel plate (MCP) as ions. 
The red points in Fig.~2(a), (b) show the obtained Ramsey interferograms at $\tau \sim 50\,$ps and $\sim 650\,$ps. 
Temporal oscillations at a period $\sim 1\,\mathrm{ femtosecond\,(fs)}$, which correspond to $E_\mathrm{eg}/\hbar \sim 2\pi \times 10^{15} \,\mathrm{Hz}$, can be clearly seen in the data.
To evaluate the effect purely arising from atomic interaction, we also measure Ramsey oscillations for a reference ``low-density'' atomic sample with mean atomic distance $n^{-1/3}\  \sim 3.5\,\mathrm{\mu m}$, which was prepared by expanding and transferring the atoms to a deep cigar-shaped optical dipole trap after the first Ramsey sequence for MI atoms. 
For each atomic preparation, we apply the Ramsey sequence for MI cloud and then utilize the remaining ground state atoms to perform the 2nd Ramsey sequence with the reference cloud. 
The Ramsey signals for the reference cloud that was recorded alternatively with those for MI are shown in the blue points in Fig.~2(a),(b).

Each Ramsey interferogram can be fitted with a function of the form $P_{\mathrm{e},i} \propto \gamma(\tau_{i})[1 + C \cos(E_\mathrm{eg} \tau_{i}/\hbar + \phi)]$
with the contrast $C$ ($C \geq 0$) and the phase $\phi$ ($-\pi \leq \phi < \pi$) as the fitting parameters. 
The prefactor $\gamma( \tau_{i} )$ that depends on the measurement index $i$ is only included for the reference cloud analysis, which takes into account the decrease in the total atom number for the 2nd Ramsey measurement due to the 1st Ramsey measurement with the MI atoms for each delay $\tau_{i}$.
The relative time-delay in each Ramsey interferogram is calibrated with attosecond precision by an optical interference signal of He-Ne laser while we scan the time-delay.
The contrasts of the Ramsey oscillations for the MI and the reference cloud are summarized in Fig.~2(c). 
The contrast for the reference cloud is constant over the whole range of the time-delay, which manifests that single-atom decoherence mechanisms can be neglected during our ultrafast dynamics. 
On the other hand, the contrast for the MI sample decreases as the time-delay increases, which we analyze in the following.
The slightly lower contrast observed in the reference cloud as compared to the MI sample near the zero-delay could be caused by small but finite inhomogeneity in the IR laser intensity over the atomic cloud~\cite{Takei2016_NCommun} and slight misalignment between the pump and the probe
beams, both of which effectively decreases the ensemble-averaged contrast for atomic samples with larger spatial size.
In the following analysis, we introduce the relative Ramsey contrast $\mathcal{C}_\mathrm{R} = C_\mathrm{H}/C_\mathrm{L}$ and the phase shift $\phi_\mathrm{R} = \phi_\mathrm{H} -\phi_\mathrm{L}$ ($-\pi \leq \phi_\mathrm{R} < \pi$), which are summarized in Fig.~3. Here, the subscript H (L) represents ``high-density'' MI atoms (``low-density'' reference atomic cloud).

\begin{figure}[t!]
    \centering
    \includegraphics[width=70mm,right]{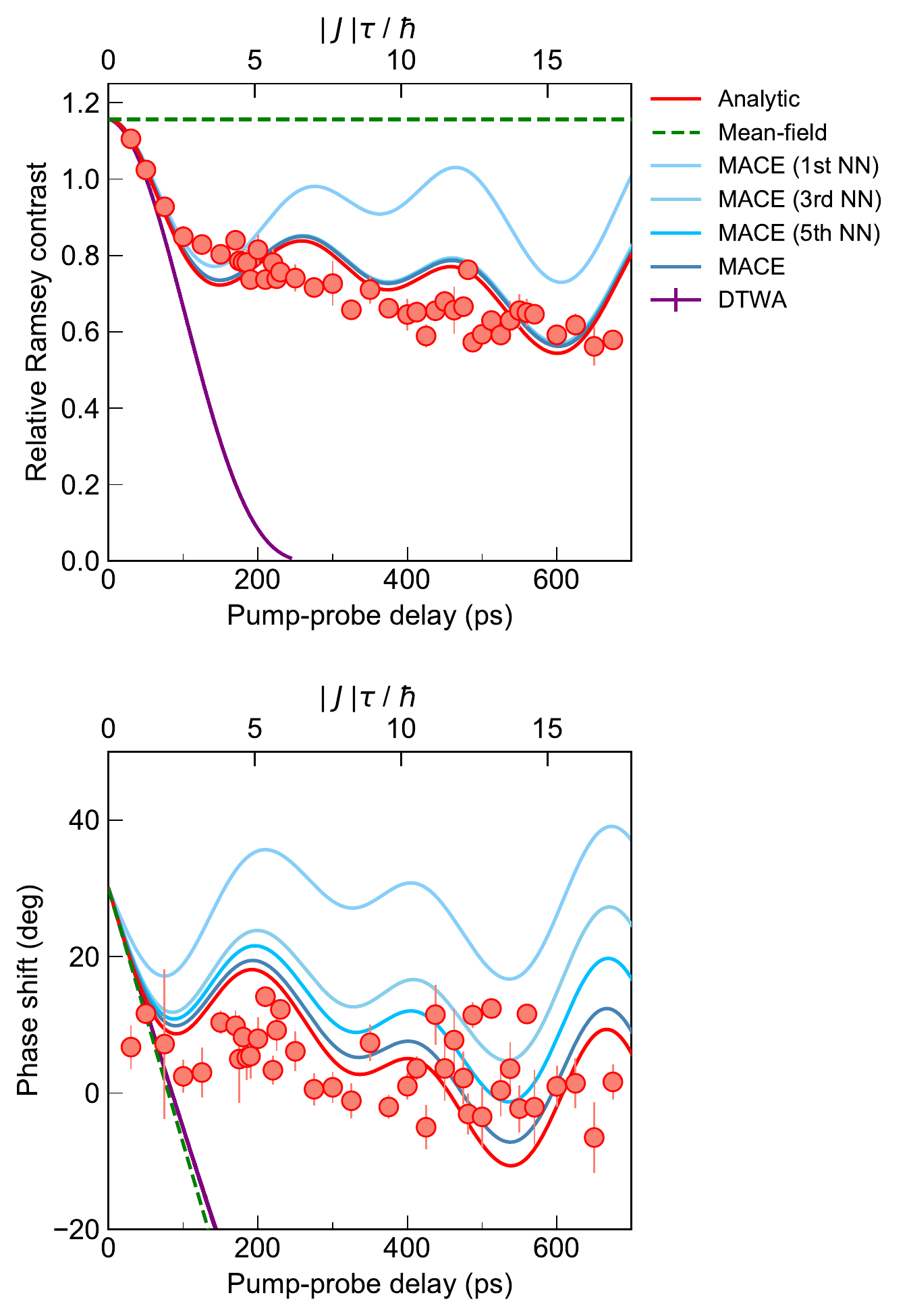}
    \caption{
    Experimental relative Ramsey contrasts and phase shifts (red circles) compared with a hierarchy of theoretical approaches. 
    The zero-delay relative Ramsey contrasts and phase shift in the theories are adjusted to have the identical value as those obtained with the analysis with the analytical solution.
    The error bars represent s.e.m.
    }
    \label{fig3}
\end{figure}

\emph{The model Hamiltonian} 
We analyze our observation by the long-range quantum Ising model~\cite{Kastner2011PRL, FossFeig2013PRA, Hazzard2014PRA, Sommer2016_PRA, Mukherjee2016PRA,Schultzen2021glassy}. 
The model Hamiltonian is
    \begin{equation}
        \hat{H} = \sum_{j} \frac{1}{2}E_\mathrm{eg} \hat{\sigma}_{j}^{z} 
        + \sum_{j<k} U_{jk} \hat{n}_j \hat{n}_k,
    \end{equation}
where the Rydberg state ($|e\rangle$) and the ground state $(|g\rangle)$ are mapped to the pseudo-spin states. 
Here $\hat{n}_j = |e \rangle \langle e|_j=(1+\hat{\sigma}_j^{z})/2$,  $\hat{\sigma}_j^{x,y,z}$ are the Pauli operators for $j$-th atom, and $U_{jk}$ is the long-range anisotropic interaction between the $j$-th and $k$-th Rydberg atoms. 
The second term in the Hamiltonian describes the effective interaction between spin-1/2 particles, which creates correlations and entanglement between distant atoms.
We consider an effective van der Waals (vdW) potential of the form $U_{jk} = -C_{6}(1-3\cos^{2}\theta_{jk})^{2}/r_{jk}^{6},$ where $C_6$ is the vdW coefficient, $r_{jk}$ is the distance between the $j$-th and $k$-th atoms, and $\theta_{jk}$ is the angle between the quantization axis and the orientation of the $j$-th and $k$-th atomic pair. 
The reason behind our assumption in the pair potential lies in the fact that calculation on the short-range pair potential below $\sim$ 1 micron is highly challenging due to the contribution from the increasing number of pair states as well as higher-order coupling beyond dipole-dipole interaction~\cite{ARC}.
Thus, in the present study, we model our system with the simplest anisotropic vdW potential with its $C_6$ as a free parameter to capture the essential many-body electron dynamics.

\emph{Exact analytical solution}
The Ising spin model in Eq.~(2) has an exact analytical solution that fully takes into account the many-body correlations~\cite{Kastner2011PRL,FossFeig2013PRA,Hazzard2014PRA,Sommer2016_PRA,Mukherjee2016PRA,Schultzen2021glassy}. 
The Ramsey signal of the $j$-th atoms is 
    \begin{equation}
        P_{\mathrm e,j}(\tau) = 2p_\mathrm{e} p_\mathrm{g} \mathrm{Re}[1+G_j(\tau)\exp[i (E_\mathrm{eg}\tau/\hbar + \varphi_{j})]],
    \end{equation}
where $G_j(\tau) = \Pi_{k \neq j}^N p_\mathrm{g} + p_\mathrm{e} \exp(i U_{jk}\tau/\hbar)$ and $\varphi_j$ is the phase acquired during the pulse excitation due to the AC Stark shift~\cite{Takei2016_NCommun}.
The single-particle function $G_j(\tau)$ contains the information about the many-body effect encoded in the Ramsey signal and related to the Larmor precession of the $j$-th pseudo-spin via
$\langle \hat{\sigma}_j^{+} (\tau)\rangle = G_j(\tau) e^{i E_\mathrm{eg} \tau/\hbar} \langle \hat{\sigma}_j^{+}(0)\rangle$, where $\hat{\sigma}_j^{+} = \hat{\sigma}_j^x +i \hat{\sigma}_j^y$.
The Ramsey contrast of the $j$-th atom $\mathcal{C}_{\mathrm{R},j}(\tau)$ is obtained by $| G_{j}(\tau) |$, and the relative phase shift of the $j$-th atom $\phi_{R,j}(\tau)$ is obtained by ${\arg}\lbrack G_{j}(\tau) \rbrack$.
The total signal is obtained by taking the average of the ensemble, $\bar{P}_\mathrm{e}(\tau) = (1/N)\sum_{j = 1}^{N} P_{\mathrm{e},j}(\tau)$. 
The pump pulse excitation creates a superposition of many-particle quantum states with a different number and arrangement of Rydberg atoms on a lattice.
During the time-delay, each many-particle state acquires a phase according to the long-range interaction and is superposed coherently to eventually give rise to the strongly-correlated many-body quantum state.

We perform a least-squares fit to the experimental contrast data with the above solution. 
Since our atomic system is large, nearly defect-free, and thus essentially homogeneous over the sample, our dynamics is well-approximated near the thermodynamic limit of $N \rightarrow \infty$, where the finite-size effect is neglected.
Using $C_{6}$ as the only free parameter, we obtain good agreement with the experimental contrast data at $C_{6}/\hbar = 371(6)\,\mathrm{MHz}\,\mathrm{\mu m^{6}}$ as shown in the red curve in Fig.~3.
Here, we forced the theory curve to pass the data point at the shortest time-delay near $\tau \sim 0$.
The angle-averaged vdW coefficient obtained from the fitted $C_6$ agrees within a factor of 3 with a numerical calculation valid at a large atomic distance~\cite{ARC}.
The $C_6$ value extracted from the contrast decay can also consistently account for the slow decreasing trend in the phase shift as shown in Fig.~3, where the zero-delay phase shift was used as a fitting parameter.
The physical mechanism behind the non-zero zero-delay phase shift is the difference in AC-Stark shifts between the two samples~\cite{SM_Mizoguchi2020_PRL}.
To gain further insights on the role of quantum correlations in our ultrafast dynamics, we compare our result with three approximation theories. 

\emph{Mean-field theory}
First, we analyze our observation with the mean-field approximation.
In the mean-field theory, the interaction energy shift of a Rydberg state is described by the sum of the interaction with the rest of the surrounding $N-1$ Rydberg atoms, $U_{\mathrm{MF},j} = \sum_{k=1, k \neq j}^{N} p_\mathrm{e} U_{jk}$.
The mean-field approximation predicts that the relative Ramsey contrast $\mathcal{C}_\mathrm{R}(\tau) = 1$, independent of $C_{6}$ and $\tau$, and the phase shift $\phi_\mathrm{R}(\tau) = U_{\mathrm{MF},j}\tau/\hbar$, which evolves linearly with $\tau$.
Similar to the above discussion, the mean-field dynamics can also be well-captured near the thermodynamic limit of $N \rightarrow \infty$.
The green dashed lines in Fig.~3 show the predicted relative Ramsey contrast and phase shift using the extracted $C_6$, both of which completely fail to account for our experimental result.
The invalidity in the mean-field prediction is pronounced in our nearly-defect free atomic array with a large number of atoms than in disorder atomic ensembles, including non-unity filled atoms in an optical lattice and small atomic arrays~\cite{Yan2013_RbK}, 
since inhomogeneity in the mean-field energy shift leads to decay and non-linear time-evolution in the ensemble-averaged contrast and phase shift, respectively.

\emph{Semi-classical theory}
The invalidity in the mean-field approach strongly suggests that quantum correlations take an essential role in our dynamics. We consider an approximation theory that can take into account quantum correlations.
The discrete truncated Wigner approximation (DTWA) is a semi-classical approach, in which the quantum uncertainty in the spin state is incorporated by the initial sampling with the discrete Wigner function~\cite{Schachenmayer2015_dTWA,Pucci2016_dTWA}.
The DTWA includes quantum corrections up to their leading-order to the mean-field theory and has been successfully explained quantum spin dynamics in various studies~\cite{Lepoutre2019,Signoles2021_3dGlassy,Orioli2018_spin_relaxation,Patscheider2020_ErXXZ}.
The comparison between our experimental result and the DTWA simulation performed with a $31^3 (\sim 3\times 10^4$)-site cubic atomic array is shown in the purple solid curves in Fig.~3.
The agreement can be confirmed only up to $\tau\sim 2\hbar/|J|$, where $J \sim -1.6\ C_6/a_\mathrm{lat}^6$ is the strongest NN interaction strength. 
The DTWA introduces initial quantum fluctuation of the spin state so that it can include quantum corrections up to their leading-order to a mean-field approximation.
Accordingly, the timescale on which the DTWA is quantitatively valid is generally set by $\mathcal {O}(\hbar/|J|)$~\cite{Sundar2019_dTWA, Kunimi2021_dTWA}, where the initial quantum fluctuation dominates the dynamics. 
For longer timescales, DTWA fails, which can be attributed to the population $p_\mathrm{e}$ being much smaller than 50\% in our setting. 
For $p_\mathrm{e} \sim 50 \%$, as in previous experiments on the many-body spin dynamics~\cite{Signoles2021_3dGlassy,Orioli2018_spin_relaxation}, DTWA accidentally coincides with the exact solution~\cite{Schachenmayer2015_dTWA}.
However, to explain the dynamics in our system, quantum effects beyond the initial quantum fluctuations must be included.

\emph{Cluster expansion theory}
Finally, we compare our result with a moving-average cluster expansion theory (MACE)~\cite{Hazzard2014_MACE}, in which the dynamics of each spin is evaluated within a cluster, whose size is increased until the dynamics converge.
The dark-blue curves in Fig.~3 show the MACE prediction for $\sim3\times{10}^4$ cubic lattice spins, which faithfully capture the observed dynamics.
The MACE simulation on cluster size dependence emulate long-range interaction with different truncation distances.
We have shown the results with 1st, 3rd, and 5th NN in Fig.~3.
The qualitatively different trend between our observation and the result with the 1st NN distance indicates that the spreading of spin-correlations is finite, but reaches beyond the NN distance within $\sim 1$ ns, which is consistent with the interaction strengths.
Interestingly, the phase shift is found to converge much slower than the contrast, indicating that our phase measurement could be a sensitive probe to detect the build-up of the long-range correlations.

In conclusion, we have developed a quantum many-body electron simulation platform that operates on the picosecond-scale by utilizing an ultrafast technique and atomic Mott insulator as a large-scale atomic array.
By comparing our observation with the solution with full many-body correlations as well as the mean-field, semi-classical and cluster expansion theories, we identify the emergence of long-range many-body correlations in our ultrafast dynamics.
Noteworthy, our unambiguous experimental identification of the essential role on many-body correlations is strongly supported by the measurability not only on the contrast decay in the Ramsey signal but also on its minute phase shift on the attosecond timescale.

The fair agreement between our observation and the analytical solution or MACE supports our treatment of the effective pair potential as long as many-body quantum correlations are properly treated in the approximation.
The quantitative difference between the experiment and the many-body simulation could be arising from the approximation to describe the Rydberg pair potential as a single effective potential as well as neglecting the finite width of the quantum ground-state atomic wavefunction in each lattice site. 
The latter could lead to uncertainty in the interaction energies between atoms.
These two factors, which are beyond the scope of our analysis, could smear out the small temporal undulations in the Ramsey oscillations at a period of a few hundreds of picoseconds expected in the many-body correlation theories (see red and blue curves in Fig.~3).

Our work presented here can be directly extended to other interaction regimes by tuning the principal quantum number $\nu$ and the orbital angular momentum (such as $S$ or $D$ orbital). Other quantum spin models such as the Heisenberg model~\cite{Nguyen2018_PRX, Signoles2021_3dGlassy}, \emph{XY} model~\cite{Orioli2018_spin_relaxation} could be implemented by employing direct ultrafast laser excitation to two different Rydberg states, which encode the spin.
Our ultrafast approach can be combined with a microscope~\cite{Zeiher2016_2dIsing, Zeiher2017_1dDressedIsing, Schauss_2dIsingCrystal, GuardadoSanchez2018PRX} as well as an optical tweezer array~\cite{deLeseleuc_SPTphase, Bluvstein2021_Scar, Omran2019_Cat, Bernien2017_51simulator, Lienhard2018_PRX,Ebadi2021_256Ryd,Scholl2021_196Ryd, Chew2021_Forster} to reveal its ultrafast dynamics with single-site resolution.
By taking advantage of our ultrafast approach, the long-time dynamics, orders of magnitude longer than the interaction timescale, could also address many-body thermalization and localization problems in our isolated quantum system~\cite{Turner2018_MBScar,Turner2018_Scar, RevModPhys2019_MBL} without being affected by environmental noises or the radiative lifetime.
Most interestingly, our ultrafast Ramsey measurement could uncover the many-body electronic states and the non-equilibrium dynamics in the metal-like quantum gas regime, where the electric charges overlap between the NN lattice site, which is an exotic state of matter that was recently realized for the first time by our ultrafast broadband laser excitation that circumvents the Rydberg blockade~\cite{Ohmori2014, Mizoguchi2020_PRL}.

The authors acknowledge Hisashi Chiba, Yasuaki Okano and Vikas Singh Chauhan for their support.
This work was supported by MEXT Quantum Leap Flagship Program (MEXT Q-LEAP) JPMXS0118069021 and JSPS Grant-in-Aid for Specially Promoted Research Grant No. 16H06289.
M.K. also acknowledges support from JSPS KAKENHI Grant No. JP20K14389.
T.F. and M.W. are supported by the Deutsche Forschungsgemeinschaft (DFG, German Research Foundation) under Germany's Excellence Strategy EXC2181/1-390900948 (the Heidelberg STRUCTURES Excellence Cluster), within the Collaborative Research Center SFB1225 (ISOQUANT) and the DFG Priority Program 1929 ``GiRyd''(DFG WE2661/12-1). T.F and M.W. acknowledge support by the European Commission FET flagship project PASQuanS (Grant No. 817482) and by the Heidelberg Center for Quantum Dynamics.

%

\end{document}